# DEVIATION OF THE WAVES IN AN INHOMOGENEOUS MEDIUM

I. SIMACIU [1,2], Z. BORSOS [1], GH. DUMITRESCU [2]

[1] Petroleum-Gas University of Ploiești, Ploiești 100680, Romania
E-mails: isimaciu@yahoo.com; borzolh@upg-ploiesti.ro
[2] High School Toma N. Socolescu, Ploiești, Romania

*Abstract.* Using the formula found by Noorbala and Sepehrinia, the wave deviation in an inhomogeneous medium with continuous variation of propagation velocity is deduced. For electromagnetic waves (light) that propagate in the gravitational field, the deduced deviation is identical to that calculated from General Relativity. The method and their consequences are a good pedagogical example that verifies the Noorbala-Sepehrinia's formula as well as the mecano-optics analogy (Hamilton's principle/ principle of stationary action and Fermat's principle) for the bodies movement in the gravitational field.

## INTRODUCTION

In a paper [1], Noorbala and Sepehrinia (N-S) found a formula which relate the refractive index $n$ to the angle $\theta$ of incidence (the angle between incident light ray and normal at the constant index surface) for the case when the speed of the light varies continuously within a medium. This new relation is different to that one of Snell' law.

We will use it in our paper to compute the deviation of the wave (particularly, the light) when it travels an inhomogeneous medium [2-4]. For this kind of medium we will assume that the refractive index depends only on the radius, $n(r)$. Our result is like that one obtained in General Relativity for an isotropic metrics [6-8].

All our derivations which we will perform in the following rows may be an opportunity to make a pedagogical practice to use the analogy between mechanics and optics, in order to prove the N-S relation and to study the motion of a body in a gravitational field [4, 5, 11].

This paper may be also a tool to improve the analytical skills of the students in order to propose mathematical models.

## N-S RELATION FOR A MEDIUM WITH REFRACTIVE INDEX $n(r)$

Let's emphasize an inhomogeneous medium with refractive index depending on the radius as

$$n(r) = \exp\left(\frac{N}{r^p}\right), N > 0, p = \frac{1}{2}, 1, \frac{3}{2}, 2, ..., . \quad (1)$$

The above dependency is suggested by Rastall-Yilmaz-Rosen metrics [12 - 15]. We will use it since it allows us to obtain an approach for the bending of the waves when the refractive index is of the form

$$n(r) = \exp\left(\frac{N}{r^p}\right) \simeq 1 + \frac{N}{r^p}, N > 0, \frac{N}{r^p} \ll 1. \quad (2)$$

For $p = 1$ and $N = r_g = 2Gm/c^2$ (where $m$ is the mass of a body placed in the origin of the reference frame), equation (2) becomes

$$n(r) = 1 + \frac{r_g}{r}. \quad (3)$$

This gravitational index of refraction is compatible with the Schwarzschild metric [9, Subch. 123].

According to the equation (10) of the paper of Noorbala and Sepehrinia [1], for continuous and inhomogeneous medium the law of refraction is

$$\sin\theta \frac{d(n\sin\theta)}{ds} = -n\cos\theta\left(\frac{d\hat{n}}{ds}\hat{\upsilon}\right). \quad (4)$$

The surface of a medium where the refractive index is constant is a spherically one.

As shown in figure 1, $\vec{r}(r,\varphi)$ is the position vector in polar coordinates, $\hat{n}$ is the normal to the surface, $\hat{\upsilon}$ is the direction of the vector $\vec{\upsilon} = d\vec{r}/ds$, $\theta$ is the angle between the vectors $\hat{n}$ and $\hat{\upsilon}$ and $\delta$ is the bending angle.

Deviation of the waves in an inhomogeneous medium

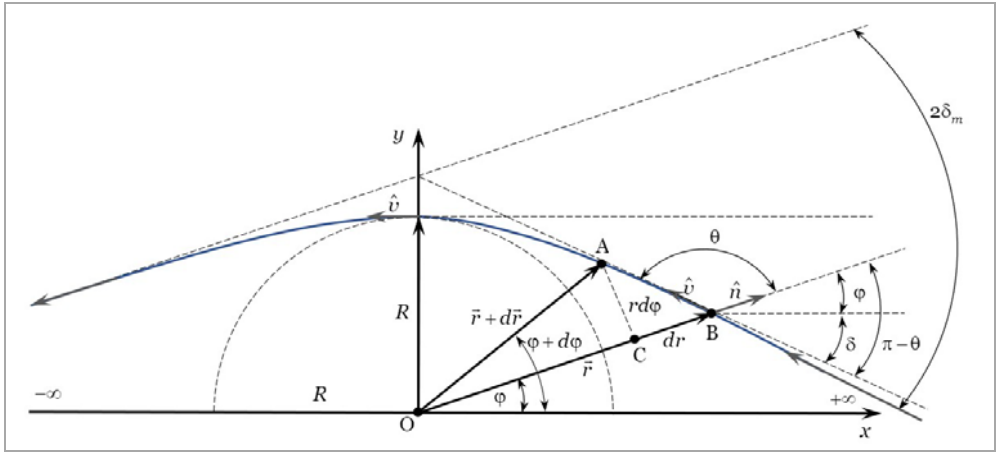

Figure 1.

Since the refractive index depends only on the length of the position vector, according to equation (2), the normal to the surface $\hat{n}$, the position vector and the gradient of the index $\nabla n = (\partial n/\partial r)(\vec{r}/r) = (\partial n/\partial r)\hat{n}$ are in a parallel direction. It follows

$$\left|\frac{d\hat{n}}{ds}\right| = \left|\frac{d}{ds}\left(\frac{\vec{r}}{r}\right)\right| = \frac{d\varphi}{ds}, \quad (5)$$

where $\varphi$ is, according to figure 1, the angular variable in polar coordinates $(r, \varphi)$ in the plane of the wave path

The vector $d\hat{n}/ds$ is perpendicular to $\hat{n} = \vec{r}/r$ and

$$\frac{d\hat{n}}{ds}\hat{v} = \frac{d\varphi}{ds}\cos\left(\frac{\pi}{2} - \theta\right) = \frac{d\varphi}{ds}\sin\theta. \quad (6)$$

According to figure 1, to figure 2 and to the notations adopted in [2] and [1] one can establish the following correspondences

$$dl \leftrightarrow ds = \sqrt{dr^2 + r^2 d\varphi^2}, \quad \alpha \leftrightarrow \theta, \quad (7)$$

When substitute equation (6) in equation (4), it follows

$$\sin\theta \frac{d(n\sin\theta)}{ds} = -n\frac{d\varphi}{ds}\cos\theta\sin\theta, \quad (8a)$$

or

$$d(n\sin\theta) = -n\cos\theta\, d\varphi, \quad (8b)$$

which is the N-S relation mentioned in the rows above.

### N-S RELATION AND THE LAW OF CONSERVATION OF ANGULAR MOMENTUM

In what it follows we will derive the law of conservation of angular momentum and we will prove the compatibility of it with N-S relation for the case when the refractive index depends on the length of the position vector.

The path of the light has the constant [2-4]. According to the equation (1) of the paper [2], this constant is

$$L = rn(r)\sin\theta. \qquad (9)$$

This constant corresponds to the conservation of the length of the angular momentum of a photon $\vec{L} = \vec{r} \times \vec{p}$ (see Appendix 1!).

At $\theta = \pi/2$ the radius becomes $r(\pi/2) = R$. Using this radius and equations (1) and (9), the constant becomes

$$L = Rn(R) = R\exp\left(\frac{N}{R^p}\right). \qquad (10)$$

One can differentiate (9), and so

$$d(n\sin\theta) = -\frac{dr}{r}n\sin\theta. \qquad (11)$$

We will use the figure 1 of this paper and the figure 2 from [2] to establish the following relations

$$\sin\theta = \frac{rd\varphi}{ds}, \quad \cos\theta = \frac{dr}{ds}. \qquad (12)$$

When substituting relations (12) within the right side of (11) one can obtain the N-S relation (8b). It is not surprisingly that these two relations are compatible.

The N-S relation was derived tacking account of the generalized Fermat's principle [1] which is in fact the Hamilton's principle for light [3, 4]. Using the Hamilton's principle one derives the angular momentum conservation [5].

### THE BENDING OF THE LIGHT TRAVELING IN AN INHOMOGENEOUS MEDIUM

We will apply N-S relation for computing the deviation of ray of light which is parallel to $Ox$ and start from a point of coordinates $r \to +\infty, \varphi \to 0$ an it is directed toward the point of coordinates $r = R, \varphi = \pi/2$ (see figure 1!). Here $R$ is the minimum length of the position vector related to the reference frame. Here the bending angle is zero, $\delta = 0$, and $\theta = \pi/2$, since these two angles are related by

$$\varphi + \delta + \theta = \pi \tag{13a}$$

and

$$d\theta = -d\varphi - d\delta. \tag{13b}$$

When express equation (7b) using $r, n$ and $\varphi$ (Appendix 2, equation A2.3), then

$$d\delta = \frac{r}{n}\frac{dn}{dr}d\varphi. \tag{14}$$

Differentiating equation (1), yields

$$\frac{dn}{dr} = -\frac{pN}{r^{p+1}}n. \tag{15}$$

Replacing equation (15) in equation (14), one obtains

$$d\delta = -\frac{pN}{r^p}d\varphi. \tag{16}$$

To integrate equation (16) it is necessary to find out how $\varphi$ depends on $r$. According to [2-4], the path of the light in a medium is depicted by (equation 2 from [2])

$$d\varphi = \pm\frac{L dr}{r\sqrt{n^2 r^2 - L^2}}, \tag{17}$$

where $L$ is the constant from equations (9) and (10).

Then we will do two replacements: first equations (1) and (10) in equation (17)

$$d\varphi = \pm\frac{R\exp(N/R^p)dr}{r\sqrt{r^2\exp(2N/r^p) - R^2\exp(2N/R^p)}} \tag{18}$$

and second equation (18) in equation (16)

$$d\delta = \mp\frac{pNR\exp(N/R^p)dr}{r^{p+1}\sqrt{r^2\exp(2N/r^p) - R^2\exp(2N/R^p)}}. \tag{19}$$

Integrating equation (19) between $r \to +\infty$ and $r = R$ we obtain the maximum deviation $\delta_m = \delta(r \to +\infty)$

$$\int_{\delta_m}^{0} d\delta = \mp \int_{+\infty}^{R} \frac{pNR\exp(N/R^p)dr}{r^{p+1}\sqrt{r^2\exp(2N/r^p) - R^2\exp(2N/R^p)}}. \quad (20a)$$

Here the integral (20a) can be solved in second order approximation for $Nr^{-p} \ll 1$

$$\int_{\delta_m}^{0} d\delta = \mp \int_{+\infty}^{R} \left( \frac{pN}{r^{p+1}\sqrt{r^2/R^2 - 1}} - \frac{pN^2 R^{-p}(r^p - R^p)}{r^{2p-1}\sqrt{r^2/R^2 - 1}(R^2 - r^2)} \right) dr. \quad (20b)$$

The solutions are for $p = 2k$

$$\delta_m = \mp \frac{\sqrt{\pi}k}{R^{4k}} N \left( -\frac{\Gamma\left(k+\frac{1}{2}\right)(R^{2k} - 2kN)}{\Gamma(k+1)} - \frac{2N\Gamma\left(2k+\frac{1}{2}\right)}{\Gamma(2k)} \right). \quad (21a)$$

The solutions are for $p = 2k + 1$

$$\delta_m = \mp \frac{\sqrt{\pi}N(2k+1)}{2R^{4k+2}} \left( \frac{\Gamma(k+1)(-R^{2k+1} + N(2k+1))}{\Gamma\left(k+\frac{3}{2}\right)} - \frac{2N\Gamma\left(2k+\frac{3}{2}\right)}{\Gamma(2k+1)} \right). \quad (21b)$$

For $p = 1 (k = 0)$

$$\delta_m = \mp \left( -\frac{N}{R} + \frac{N^2}{R^2} - \frac{\pi N^2}{2R^2} \right) \quad (21c)$$

is an analytical formula using WolframAlpha or Wolfram Programing Lab [16].

### THE DEVIATION OF THE LIGHT IN A GRAVITATIONAL FIELD

Assuming a gravitational field of a body with mass $m$, the refractive index can be expressed like (3), for the first-order isotropic Schwarzshild metric [6-8]. Therefore, the maximum deviation, with $p = 1$ and $N = r_g$ in equation (21c), [16], is

$$\delta_m \cong \pm \frac{r_g}{R}. \quad (22)$$

The entire deviation, which occurs when the wave travels from $r \to +\infty$ to $r \to -\infty$, and passing at least distance $R$, is

Deviation of the waves in an inhomogeneous medium

$$\delta_g = 2\delta_m = \frac{2r_g}{R}. \qquad (23)$$

This result is just the relativistic deviation [6, 9]. That is, an inhomogeneous optic medium with a refractive index of the form (3) behaves like a spherical lens and this lens mimics a gravitational field.

One can rich the same result using lensmaker's equation (36) from subchapter 4.4. of the paper [3] for a spherical lens with radii of curvature $R_1 = R_2 = R$ and refractive index $n_p(R) = 1 + N/R^p$

$$\frac{1}{f_p} = [n(R) - 1]\frac{2}{R} = \frac{2N}{R^{p+1}}. \qquad (24)$$

Then, the entire deviation for a parallel ray to the direction and which pass through a point of coordinates $r = R$ is

$$\delta_p \simeq \tan\delta_p = \frac{R}{f_p} = \frac{2N}{R^p}. \qquad (25)$$

For $p = 1$ and $N = r_g$ the entire deviation is

$$\delta_1 = \delta_g = \frac{R}{f_g} = \frac{2r_g}{R}. \qquad (26)$$

and this is just the relativistic deviation from equation (23).

**CONCLUSIONS**

According to the general relativity, half of the bending of the light in a gravitational field depends on the curvature of the space and the other half to the variation of the velocity of the light along its path [9].

An optical approach has to assume that the deviation of the waves, no matter of their kind [10], and for an inhomogeneous medium, $n(r)$, depends on the optical properties of the medium, as we shown in this paper.

Since in the classic physics the bending of the path it is assumed to the action of a force, then a gravitational force (i.e. which is directly proportional to the mass and inversely proportional to the distance) is also the effect of a refractive index which depends on the position vector. Therefore, if will be able to assume a scenario of becoming inhomogeneous the medium around a particle, then we will be able to get a phenomenological-causal approach of the gravitational interaction into the electromagnetic world, i.e. the world where the main interaction is the electromagnetic one.

For a refractive index from equation (3), the acceleration is directly proportional to $r_g/r^2$, i.e. a gravitational type of acceleration.

This kind of approach had succeed to apply the mechano-optics of the gravitational interaction to the research of the light and also to the particles with rest mass [11].

**Appendix 1**

According to the mechano-optics the angular momentum of a photon which moves in a gravitational field is

$$\left|\vec{L}_m\right| = \left|\vec{r} \times \vec{p}\right| = rp\sin\theta. \tag{A1.1}$$

Its momentum is

$$p = m\upsilon = \hbar k. \tag{A1.2}$$

where $k$ is the wave number.

Using the angular speed $\omega$, we may express the mass of the photon as $m = \hbar\omega/c$, the speed as $\upsilon = c/n$ and the momentum as

$$p = \frac{\hbar\omega}{c}\upsilon = \hbar k = \hbar\frac{n\omega}{c}. \tag{A1.3}$$

Replacing the equations (A1.3) and (A1.2) in equation (A1.1), yields

$$\left|\vec{L}_m\right| = \left(\frac{\hbar\omega}{c}\right)rn\sin\theta = \left(\frac{\hbar\omega}{c}\right)L. \tag{A1.4}$$

Since the angular momentum is a constant during its motion and also $\hbar\omega/c$ (the angular speed doesn't change during the travelling through the medium), then the following parameter

$$L = \frac{c\left|\vec{L}_m\right|}{\hbar\omega} = rn(r)\sin\theta \tag{A1.5}$$

is a constant.

**Appendix 2**

Equation (8b) can be put also in the form

$$dn\sin\theta + n\cos\theta d\theta = -nd\varphi\cos\theta. \tag{A2.1}$$

Replacing $d\theta$ from equation (13b) in equation (A2.1), one obtains

Deviation of the waves in an inhomogeneous medium

$$d\delta = \frac{dn}{n}\frac{\sin\theta}{\cos\theta}. \qquad (A2.2)$$

Finally using equation (12) in equation (A2.2), it follows

$$d\delta = \frac{r}{n}\frac{dn}{dr}d\varphi. \qquad (A2.3)$$